\documentclass{article}
\usepackage[subtle]{savetrees}
\usepackage{spconf,amsmath,graphicx,hyperref,booktabs,tabularx, multirow, amsfonts, bm, cite}
\DeclareMathOperator*{\argmin}{arg\,min}
\usepackage{kotex}
\usepackage{makecell}
\usepackage{threeparttable}
\hypersetup{ 
    colorlinks=true,
    linkcolor=purple,
    filecolor=magenta,      
    urlcolor=cyan,
}

\usepackage[table]{xcolor}
\usepackage{array,colortbl}
\definecolor{TRKDhl}{RGB}{255,250,205}   
\definecolor{deltag}{RGB}{230,250,230}   
\definecolor{deltar}{RGB}{250,230,230}   
\newcolumntype{C}{>{\centering\arraybackslash}m{2.4em}} 
\newcolumntype{G}{>{\centering\arraybackslash}m{0.6em}} 
\newcommand{\deltaup}[1]{\cellcolor{deltag}{#1}}
\newcommand{\deltadown}[1]{\cellcolor{deltar}{#1}}

\title{Triage knowledge distillation for speaker verification}
\name{Ju-ho Kim, Youngmoon Jung, Joon-Young Yang, Jaeyoung Roh, Chang Woo Han, Hoon-Young Cho}
\address{AI Solution Team, Samsung Research, Seoul, South Korea}

\begin{document}
\ninept
\maketitle

\begin{abstract}
Deploying speaker verification on resource-constrained devices remains challenging due to the computational cost of high-capacity models; knowledge distillation (KD) offers a remedy. 
Classical KD entangles target confidence with non-target structure in a Kullback--Leibler term, limiting the transfer of relational information.
Decoupled KD separates these signals into target and non-target terms, yet treats non-targets uniformly and remains vulnerable to the long tail of low-probability classes in large-class settings. 
We introduce Triage KD (TRKD), a distillation scheme that operationalizes \textit{assess--prioritize--focus}.
TRKD introduces a cumulative-probability cutoff $\tau$ to \textit{assess} per-example difficulty and partition the teacher posterior into three groups: the target class, a high-probability non-target confusion-set, and a background-set.
To \textit{prioritize} informative signals, TRKD distills the confusion-set conditional distribution and discards the background.
Concurrently, it transfers a three-mass (target/confusion/background) that capture sample difficulty and inter-class confusion. 
Finally, TRKD \textit{focuses} learning via a curriculum on $\tau$: training begins with a larger $\tau$ to convey broad non-target context, then $\tau$ is progressively decreased to shrink the confusion-set, concentrating supervision on the most confusable classes. 
In extensive experiments on VoxCeleb1 with both homogeneous and heterogeneous teacher--student pairs, TRKD was consistently superior to recent KD variants and attained the lowest EER across all protocols. 
\end{abstract}


\section{Introduction}
\label{sec:intro}
Recent advances in speaker verification (SV) have been driven by high-capacity models trained on large-scale datasets, but their computational and memory footprints hinder real-time, on-device deployment \cite{chen2022large, chen2022wavlm}. 
Knowledge distillation (KD) \cite{hinton2015distilling} offers a practical remedy: a student network is trained to match a teacher network’s soft posterior, thereby inheriting the teacher’s generalization and ``dark knowledge'' that encodes fine-grained inter-class similarities and decision-boundary structure \cite{lee2022fithubert, heo2023one}.

Classical KD optimizes a single Kullback--Leibler (KL) divergence between teacher and student posteriors, entangling target and non-target supervision in a single term \cite{hinton2015distilling}. 
When the teacher is highly confident in the target class, this coupling suppresses informative non-target structure \cite{zhao2022decoupled, truong2024emphasized}. 
Decoupled KD (DKD) \cite{zhao2022decoupled} addresses this by decomposing the objective into target-class KD (TCKD), which conveys per-example difficulty via the teacher’s target confidence, and non-target-class KD (NCKD), which transfers relational structure among the remaining classes. 

Although DKD improves inter-class similarity transfer, scaling to large-class identification brings additional challenges. 
A teacher’s probability mass typically concentrates on the target and a few confusable impostors, leaving a long tail of near-zero probabilities \cite{zhao2023grouped, gan2025grouped}. 
Aligning the full non-target distribution dilutes supervision: individually negligible classes carry little information, yet their accumulated gradients impede optimization. 
Grouped KD (GKD) \cite{zhao2023grouped} mitigates this by transferring only a high-probability subset of the teacher's distribution. 
However, with a static cutoff (e.g., top-k or a fixed probability threshold), enforcing fine-grained alignment too early can overwhelm the student and destabilize training.

We propose \textbf{Triage KD (TRKD)}, which instantiates the \textit{assess--prioritize--focus} triage principle within the distillation objective, building on difficulty-aware curricula and selective supervision \cite{bengio2009curriculum, shrivastava2016training, lin2017focal}. 
First, we \textit{assess} class-wise difficulty using a cumulative-probability cutoff $\tau$. 
At each training step, the teacher posterior is partitioned into three groups: the target class, a confusion-set formed by the smallest collection of non-targets whose cumulative probability exceeds $\tau$, and a low-probability background.  
Second, to \textit{prioritize} informative signals, TRKD aligns the conditional distribution over the confusion-set while discarding the background. 
In parallel, it transfers a three-mass (target/confusion/background) that encodes sample difficulty and inter-class confusion. 
Finally, TRKD \textit{focuses} learning through a curriculum on $\tau$: training starts with a larger $\tau$ (broad non-target context) and $\tau$ is gradually decreased to shrink the confusion-set, concentrating supervision on the hardest impostors. 
This schedule stabilizes the student's optimization in early training and sharpens decision boundaries later on, without requiring any architectural changes or additional data. 

We validated TRKD across diverse teacher--student combinations (heterogeneous and homogeneous). 
All models were trained on the VoxCeleb2 development set and evaluated on standard VoxCeleb1 trials. 
TRKD consistently outperformed recent logit- and embedding-based KD methods across all protocols, and ablations attributed the gains to (i) three-mass partition-and-transfer, (ii) explicit confusion-set alignment, and (iii) the cumulative-probability cutoff curriculum on $\tau$.

\section{Methodology}

\subsection{Preliminaries}

\begin{figure*}[!t]
\centering
\includegraphics[width=\textwidth]{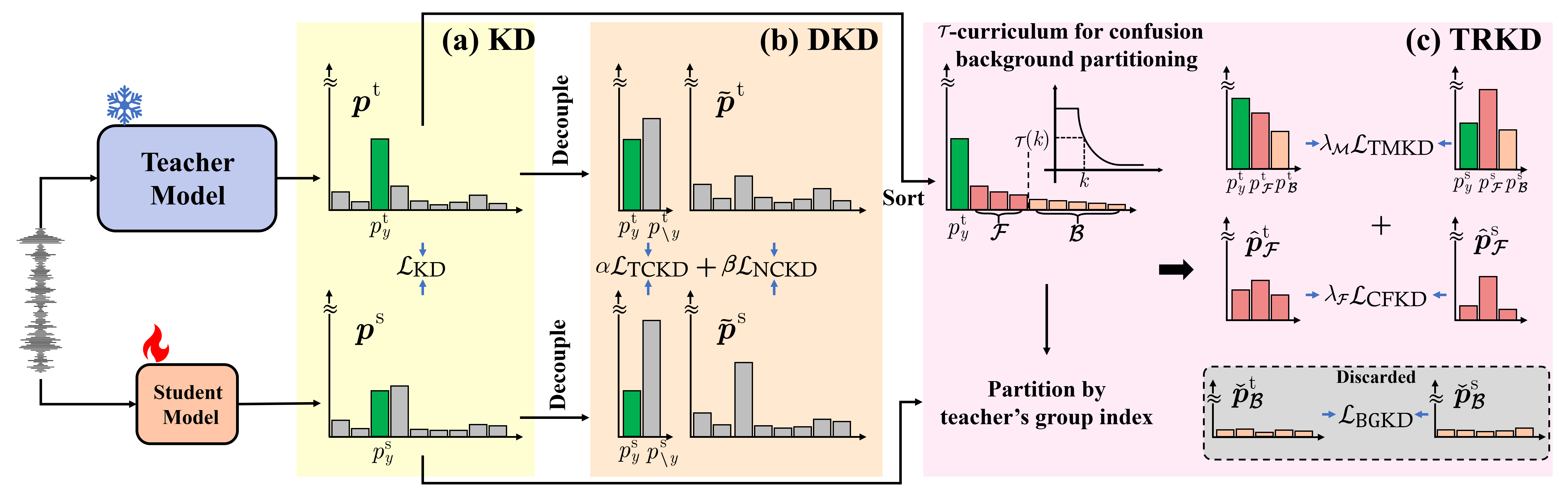}
\vspace{-5mm}
\caption{
\textbf{Comparison: KD vs.\ DKD vs.\ TRKD.}
\textbf{(a)} Classical KD aligns teacher and student posteriors via
$\mathcal{L}_{\mathrm{KD}}=\mathrm{KL}(\boldsymbol{p}^{\mathrm t}\,\|\,\boldsymbol{p}^{\mathrm s})$.
\textbf{(b)} DKD decouples KD into a target--non-target KL term, 
$\mathcal{L}_{\mathrm{TCKD}}=\mathrm{KL}([p^{\mathrm t}_y,p^{\mathrm t}_{\setminus y}]\,\|\, [p^{\mathrm s}_y,p^{\mathrm s}_{\setminus y}])$, 
and a normalized non-target KL term, 
$\mathcal{L}_{\mathrm{NCKD}}=\mathrm{KL}(\tilde{\boldsymbol{p}}^{\mathrm t}\,\|\,\tilde{\boldsymbol{p}}^{\mathrm s})$, enhancing transfer among non-target classes.
\textbf{(c)} TRKD partitions probabilities into three masses (target $y$, a high-probability non-target confusion-set $\mathcal F$, and a low-probability non-target background-set $\mathcal B$) and shrinks the confusion-set via a cumulative-probability cutoff $\tau(k)$ that decreases during training. 
The student minimizes a three-mass KL,
$\mathcal{L}_{\mathrm{TMKD}}=\mathrm{KL}([p^{\mathrm t}_y,p^{\mathrm t}_{\mathcal F},p^{\mathrm t}_{\mathcal B}]\,\|\, [p^{\mathrm s}_y,p^{\mathrm s}_{\mathcal F},p^{\mathrm s}_{\mathcal B}])$,
with a confusion-set conditional KL,
$\mathcal{L}_{\mathrm{CFKD}}=\mathrm{KL}(\hat{\boldsymbol p}^{\mathrm t}_{\mathcal F}\,\|\,\hat{\boldsymbol p}^{\mathrm s}_{\mathcal F})$;
the background-set term $\mathcal{L}_{\mathrm{BGKD}}$ is discarded to suppress long-tail noise.
}
\label{fig:TRKD_method}
\vspace{-2mm}
\end{figure*}

\paragraph*{Classical KD.}
Let $C$ be the number of training classes with index set $\mathcal{C}=\{1,\dots,C\}$. 
The teacher and student logit vectors are $\boldsymbol{z}^{\mathrm{t}},\boldsymbol{z}^{\mathrm{s}}\in\mathbb{R}^{C}$ with components $z^{\mathrm{t}}_{i},z^{\mathrm{s}}_{i}$ for $i\in\mathcal{C}$.
The corresponding class-probability vectors $\boldsymbol{p}^{\mathrm t}$ and $\boldsymbol{p}^{\mathrm s}$ are given by the softmax:
\begin{equation}
\boldsymbol{p}^{\mathrm{t}}=\mathrm{softmax}(\boldsymbol{z}^{\mathrm{t}}),\quad
\boldsymbol{p}^{\mathrm{s}}=\mathrm{softmax}(\boldsymbol{z}^{\mathrm{s}}),
\end{equation}
i.e., for each component $i\in\mathcal{C}$:
\begin{equation}
p_i^{\mathrm{t}}=\frac{\exp(z_i^{\mathrm{t}})}{\sum\limits_{j\in\mathcal C} \exp(z_j^{\mathrm{t}})},\quad
p_i^{\mathrm{s}}=\frac{\exp(z_i^{\mathrm{s}})}{\sum\limits_{j\in\mathcal C} \exp(z_j^{\mathrm{s}})}.
\label{eq:softmax}
\end{equation}
For brevity, temperature scaling is omitted in our formulation. 
Classical KD (Fig.~\ref{fig:TRKD_method}(a)) aligns the teacher and student distributions by minimizing the KL divergence between them:
\begin{equation}
\mathcal{L}_{\mathrm{KD}}
=\mathrm{KL}\!\big(\boldsymbol{p}^{\mathrm{t}}\,\big\|\,\boldsymbol{p}^{\mathrm{s}}\big)
=\sum_{i\in\mathcal{C}} p^{\mathrm{t}}_i\,\log\!\left(\frac{p^{\mathrm{t}}_i}{p^{\mathrm{s}}_i}\right).
\label{eq:kd} 
\end{equation}

\paragraph*{Decoupled KD (DKD).}
While classical KD transfers cross-class information, its single KL term entangles two signals: (i) the teacher’s confidence in the ground-truth class, and (ii) the relational structure among the remaining classes. 
When the teacher is highly confident, the former can dominate the latter. 
To address this, Zhao \textit{et al.}~\cite{zhao2022decoupled} decompose the KD loss into target and non-target terms: 
\begin{align}
\mathcal{L}_{\mathrm{KD}}
&=\sum_{i\in\mathcal C} p^{\mathrm{t}}_{i}\,\log\!\left(\frac{p^{\mathrm{t}}_{i}}{p^{\mathrm{s}}_{i}}\right)
= p^{\mathrm{t}}_{y}\log\!\left(\frac{p^{\mathrm{t}}_{y}}{p^{\mathrm{s}}_{y}}\right)
+ \sum_{i\in\mathcal{C}\setminus\{y\}} p^{\mathrm{t}}_{i}\log\!\left(\frac{p^{\mathrm{t}}_{i}}{p^{\mathrm{s}}_{i}}\right),
\label{eq:kd-split}
\end{align}
where $y$ is the ground-truth class index and $\mathcal {C} \setminus \{y\}$ denotes the non-target classes. 
Let $p^{\mathrm t}_{\setminus y}=1-p^{\mathrm t}_y$ and $p^{\mathrm s}_{\setminus y}=1-p^{\mathrm s}_y$ be the total non-target probability masses for teacher and student. 
For each $i\in\mathcal C\setminus\{y\}$, define the normalized non-target probabilities 
$\tilde p^{\mathrm t}_i = p^{\mathrm t}_i / p^{\mathrm t}_{\setminus y}$ and
$\tilde p^{\mathrm s}_i = p^{\mathrm s}_i / p^{\mathrm s}_{\setminus y}$.
Then the second term of Eq.~\eqref{eq:kd-split} expands as:
\begin{align}
\sum_{i\in\mathcal{C} \setminus \{y\}} p^{\mathrm{t}}_{i}\log\!\left(\frac{p^{\mathrm{t}}_{i}}{p^{\mathrm{s}}_{i}}\right)
= p^{\mathrm{t}}_{\setminus y} \log\!\left(\frac{p^{\mathrm{t}}_{\setminus y}}{p^{\mathrm{s}}_{\setminus y}}\right)
+ p^{\mathrm{t}}_{\setminus y} \sum_{i\in\mathcal{C} \setminus \{y\}} \tilde p^{\mathrm{t}}_{i}\log\!\left(\frac{\tilde p^{\mathrm{t}}_{i}}{\tilde p^{\mathrm{s}}_{i}}\right).
\label{eq:cond-split}
\end{align}
Consequently, the classical KD loss can be decomposed into two parts:
\begin{align}
\mathcal{L}_{\mathrm{KD}}
&=\underbrace{\mathrm{KL}\!\Big(\,[\,p^{\mathrm{t}}_{y},\,p^{\mathrm{t}}_{\setminus y}\,]\,\Big\|\, [\,p^{\mathrm{s}}_{y},\,p^{\mathrm{s}}_{\setminus y}\,]\,\Big)}_{\mathcal{L}_{\mathrm{TCKD}}}
\;+\;
\underbrace{p^{\mathrm{t}}_{\setminus y}\!\sum_{i\in\mathcal{C}\setminus\{y\}} \tilde p^{\mathrm{t}}_{i}\,
\log\!\Big(\frac{\tilde p^{\mathrm{t}}_{i}}{\tilde p^{\mathrm{s}}_{i}}\Big)}_{\text{non-target conditional term}}.
\label{eq:kd-decompose}
\end{align}
Here, the first term is a KL divergence between the target vs. non-target mass, and the second term is a weighted non-target conditional loss. 
As illustrated in Fig.~\ref{fig:TRKD_method}(b), DKD removes the sample-dependent factor $p^{\mathrm{t}}_{\setminus y}$ and assigns independent weights $\alpha$ and $\beta$ to the two terms:
\begin{equation}
\mathcal{L}_{\mathrm{DKD}}
= \alpha\,\mathcal{L}_{\mathrm{TCKD}}
+ \beta\,\mathcal{L}_{\mathrm{NCKD}},\qquad
\mathcal{L}_{\mathrm{NCKD}}
=\mathrm{KL}\!\big(\tilde{\boldsymbol{p}}^{\mathrm t}\,\big\|\,\tilde{\boldsymbol{p}}^{\mathrm s}\big).
\label{eq:dkd}
\end{equation}
In the above, $\tilde{\boldsymbol{p}}^t$ and $\tilde{\boldsymbol{p}}^s$ are the teacher and student probability vectors over the non-target classes after normalization.
By reducing the influence of the teacher’s target confidence, DKD alleviates target dominance and strengthens the transfer of non-target relational structure.

\subsection{Proposed: Triage KD (TRKD)}
While DKD mitigates target dominance by decoupling the KD loss, it still assigns uniform importance to all non-target classes. 
In large-scale classification, teacher posteriors typically concentrate on the target and a few confusable impostors, leaving a long tail of near-zero probabilities that contribute marginal supervision yet increase the optimization burden \cite{zhao2023grouped}.

We propose Triage KD (TRKD) (Fig.~\ref{fig:TRKD_method}(c)), which operationalizes \textit{assess--prioritize--focus}:
(i) \textit{assess} by partitioning the teacher posterior into three masses (target, a high-probability non-target confusion-set, and a low-probability non-target background-set);
(ii) \textit{prioritize} by aligning the student to the teacher within the confusion-set and enforcing a coarse three-mass constraint (target/confusion/background), while discarding fine-grained alignment over the background; 
and (iii) \textit{focus} learning via a curriculum on the cumulative-probability cutoff $\tau$, starting with a larger $\tau$ to transfer broad non-target context and then $\tau$ is decreased to shrink the confusion-set and concentrate supervision on the hardest impostors. 
\vspace{-1mm}

\paragraph*{$\tau$-curriculum for confusion--background partitioning.}
Let $k$ be the current training step, and let $k_{\text{start}}, k_{\text{stop}}$ denote the start and end of the scheduling phase. 
We define a clipped progress function $v(k)\in[0,1]$ as:
\begin{equation}
v(k)=\min\!\Big\{1,\max\!\Big\{0,\,\frac{k-k_{\text{start}}}{\,k_{\text{stop}}-k_{\text{start}}\,}\Big\}\Big\}.
\label{eq:clip}
\end{equation}
An exponential schedule is used to control the cutoff:
\begin{equation}
\tau(k)=
\begin{cases}
\tau_{\text{init}}, & k<k_{\text{start}},\\[2pt]
\tau_{\text{init}}+(\tau_{\text{final}}-\tau_{\text{init}})\,\big(1-\gamma^{\,v(k)}\big), & k_{\text{start}}\le k<k_{\text{stop}},\\[2pt]
\tau_{\text{final}}, & k\ge k_{\text{stop}},
\end{cases}
\label{eq:tau_schedule}
\end{equation}
with curvature parameter $\gamma\in[0,1]$.
At step $k$, we sort the non-target indices $\mathcal C\setminus\{y\}$ in descending order of $p_i^{\mathrm t}$ and define the confusion-set $\mathcal{F}_{\tau(k)}$ as the smallest subset ($\mathcal S$) whose cumulative probability is at least $\tau(k)$; the remaining non-target classes form the background-set $\mathcal B_{\tau(k)}$:
\begin{equation}
\mathcal{F}_{\tau(k)}\;=\;\argmin_{\mathcal{S}\subseteq\mathcal{C} \setminus \{y\}}
\Big\{\,\mid\,\mathcal{S}\,\mid \;\big|\; \sum_{i\in\mathcal{S}} p_i^{\mathrm t} \ge \tau(k)\Big\}, 
\quad
\mathcal{B}_{\tau(k)}\;=\;\big(\mathcal{C}\setminus\{y\}\big)\setminus \mathcal{F}_{\tau(k)}.
\label{eq:TRKD-prefix}
\end{equation}
For simplicity, we write $\mathcal{F}\equiv \mathcal{F}_{\tau(k)}$ and $\mathcal{B}\equiv \mathcal{B}_{\tau(k)}$.
Let the aggregated probabilities be
$
p^{\circ}_{\mathcal F}=\sum_{i\in\mathcal F}p^{\circ}_i
$
and
$
p^{\circ}_{\mathcal B}=\sum_{j\in\mathcal B}p^{\circ}_j 
$ for $\circ\in\{\mathrm t,\mathrm s\}$.
We define the within-set normalized distributions:
\begin{equation}
\hat p^{\circ}_{i\,\mid\,\mathcal F}=\frac{p^{\circ}_i}{p^{\circ}_{\mathcal F}}\ (i\in\mathcal F),
\qquad
\check p^{\circ}_{j\,\mid\,\mathcal B}=\frac{p^{\circ}_j}{p^{\circ}_{\mathcal B}}\ (j\in\mathcal B).
\label{eq:TRKD-masses}
\end{equation}
(Note that the total non-target mass decomposes as $p^{\circ}_{\setminus y}=p^{\circ}_{\mathcal F}+p^{\circ}_{\mathcal B}$.)

\paragraph*{Derivation from KD.}
Starting from Eq.~\eqref{eq:kd}, the KL divergence can be decomposed under the partition $\{\mathcal F,\mathcal B\}$:
\begin{equation}
\begin{split}
\mathrm{KL}(\boldsymbol{p}^{\mathrm t}\,\|\,\boldsymbol{p}^{\mathrm s})
&= p^{\mathrm{t}}_{y}\log\!\left(\frac{p^{\mathrm{t}}_{y}}{p^{\mathrm{s}}_{y}}\right)
+ \sum_{i\in\mathcal F} p^{\mathrm t}_{i}\log\!\left(\frac{p^{\mathrm t}_{i}}{p^{\mathrm s}_{i}}\right)
+ \sum_{j\in\mathcal B} p^{\mathrm t}_{j}\log\!\left(\frac{p^{\mathrm t}_{j}}{p^{\mathrm s}_{j}}\right).
\end{split}
\label{eq:TRKD-expand}
\end{equation}
Using the definitions above, each non-target summation factorizes into a mass term and a within-set conditional term:
\begin{equation}
\begin{aligned}
\sum_{i\in\mathcal F} p^{\mathrm t}_{i}\log\!\left(\frac{p^{\mathrm t}_{i}}{p^{\mathrm s}_{i}}\right)
&=  p^{\mathrm t}_{\mathcal F}\log\!\left(\frac{p^{\mathrm t}_{\mathcal F}}{p^{\mathrm s}_{\mathcal F}}\right)
 + p^{\mathrm t}_{\mathcal F}\!\sum_{i\in\mathcal F} \hat p^{\mathrm t}_{i\mid\mathcal F}
 \log\!\left(\frac{\hat p^{\mathrm t}_{i\mid\mathcal F}}{\hat p^{\mathrm s}_{i\mid\mathcal F}}\right),\\
\sum_{j\in\mathcal B} p^{\mathrm t}_{j}\log\!\left(\frac{p^{\mathrm t}_{j}}{p^{\mathrm s}_{j}}\right)
&= p^{\mathrm t}_{\mathcal B}\log\!\left(\frac{p^{\mathrm t}_{\mathcal B}}{p^{\mathrm s}_{\mathcal B}}\right)
+ p^{\mathrm t}_{\mathcal B}\!\sum_{j\in\mathcal B} \check p^{\mathrm t}_{j\mid \mathcal B}
 \log\!\left(\frac{\check p^{\mathrm t}_{j\mid \mathcal B}}{\check p^{\mathrm s}_{j\mid \mathcal B}}\right).
\end{aligned}
\label{eq:TRKD-factor}
\end{equation}
Substituting these into Eq.~\eqref{eq:TRKD-expand} yields a two-level decomposition with a three-mass term and separate conditional terms for the confusion- and background-set: 
\begin{equation}
\begin{aligned}
\mathrm{KL}(\boldsymbol{p}^{\mathrm t}\,\|\,\boldsymbol{p}^{\mathrm s})
&=\underbrace{\mathrm{KL}\!\Big(\,[p^{\mathrm t}_{y},p^{\mathrm t}_{\mathcal F},p^{\mathrm t}_{\mathcal B}]\,\Big\|\, [p^{\mathrm s}_{y},p^{\mathrm s}_{\mathcal F},p^{\mathrm s}_{\mathcal B}]\,\Big)}_{\text{three-mass KL}}\\
&\quad+\ \underbrace{p^{\mathrm t}_{\mathcal F}\ \sum_{i\in\mathcal F} \hat p^{\mathrm t}_{i\mid\mathcal F}
 \log\!\left(\frac{\hat p^{\mathrm t}_{i\mid\mathcal F}}{\hat p^{\mathrm s}_{i\mid\mathcal F}}\right)}_{\text{confusion-set conditional}}\
+\ \underbrace{p^{\mathrm t}_{\mathcal B}\ \sum_{j\in\mathcal B} \check p^{\mathrm t}_{j\mid \mathcal B}
 \log\!\left(\frac{\check p^{\mathrm t}_{j\mid \mathcal B}}{\check p^{\mathrm s}_{j\mid \mathcal B}}\right)}_{\text{background-set conditional}}.
\end{aligned}
\label{eq:TRKD-2level}
\end{equation}

\paragraph*{Objective function.}
From Eq.~\eqref{eq:TRKD-2level}, we identify three loss components (omitting the sample-dependent prefactors $p^{\mathrm t}_{\mathcal F}$ and $p^{\mathrm t}_{\mathcal B}$):
\begin{equation}
\begin{aligned}
&\mathcal{L}_{\mathrm{TMKD}}
=\mathrm{KL}\!\Big(\,[p^{\mathrm t}_{y},p^{\mathrm t}_{\mathcal F},p^{\mathrm t}_{\mathcal B}]\,\Big\|\, [p^{\mathrm s}_{y},p^{\mathrm s}_{\mathcal F},p^{\mathrm s}_{\mathcal B}]\,\Big),\\
&\mathcal{L}_{\mathrm{CFKD}}
=\mathrm{KL}\!\big(\hat{\boldsymbol{p}}^{\mathrm t}_{\mathcal F}\ \big\|\ \hat{\boldsymbol{p}}^{\mathrm s}_{\mathcal F}\big)
=\sum_{i\in\mathcal F} \hat p^{\mathrm t}_{i\mid\mathcal F}
 \log\!\left(\frac{\hat p^{\mathrm t}_{i\mid\mathcal F}}{\hat p^{\mathrm s}_{i\mid\mathcal F}}\right),\\
&\mathcal{L}_{\mathrm{BGKD}}
=\mathrm{KL}\!\big(\check{\boldsymbol{p}}^{\mathrm t}_{\mathcal B}\ \big\|\ \check{\boldsymbol{p}}^{\mathrm s}_{\mathcal B}\big)
=\sum_{j\in\mathcal B} \check p^{\mathrm t}_{j\mid \mathcal B}
 \log\!\left(\frac{\check p^{\mathrm t}_{j\mid \mathcal B}}{\check p^{\mathrm s}_{j\mid \mathcal B}}\right).
\label{eq:each-TRKD-loss}
\end{aligned}
\end{equation}
Here, for $\circ\in\{\mathrm t,\mathrm s\}$, $\hat{\boldsymbol{p}}^{\circ}_{\mathcal F}$ and $\check{\boldsymbol{p}}^{\circ}_{\mathcal B}$ are the within-set normalized distribution vectors over the confusion-set $\mathcal F$ and the background-set $\mathcal B$, respectively.
Following prior observations on long-tail noise in distillation~\cite{zhao2023grouped}, TRKD retains the three-mass term ($\mathcal{L}_{\mathrm{TMKD}}$) and the confusion-set conditional term ($\mathcal{L}_{\mathrm{CFKD}}$) while discarding the background-set term ($\mathcal{L}_{\mathrm{BGKD}}$).
Replacing the sample-dependent factors with fixed weights $\lambda_\mathcal{M}$ and $\lambda_\mathcal{F}$, we define the TRKD objective as:
\begin{equation}
\mathcal{L}_{\mathrm{TRKD}}
= \lambda_\mathcal{M}\,\mathcal{L}_{\mathrm{TMKD}}
+ \lambda_\mathcal{F}\,\mathcal{L}_{\mathrm{CFKD}}.
\label{eq:TRKD-loss}
\end{equation}
With the curriculum-based cutoff $\tau(k)$ in Eq.~\eqref{eq:tau_schedule}, early training transfers coarse non-target context, whereas later training focuses on the most confusable non-targets, filtering out long-tail noise. 
The overall loss function is the sum of the standard additive angular margin softmax (AAM-Softmax; $\mathcal{L}_{\mathrm{AAM}}$) \cite{deng2019arcface} and TRKD loss: 
\begin{equation}
\mathcal{L}
= \mathcal{L}_{\mathrm{AAM}}
+ \mathcal{L}_{\mathrm{TRKD}}.
\label{eq:overall-loss}
\end{equation}

\section{Experimental Setup}

\paragraph*{Datasets and evaluation protocol.}
Models were trained on the VoxCeleb2 development set~\cite{chung2018voxceleb2} and evaluated on the VoxCeleb1 \cite{nagrani2017voxceleb} original (O), extended (E), and hard (H) trials according to the official protocols. 
Performance was reported in terms of equal error rate (EER, \%).
\vspace{-4mm}

\paragraph*{Architectures and implementations.}
To assess generality, we explored diverse teacher--student pairs spanning ECAPA-TDNNs~\cite{desplanques2020ecapa}, ResNets~\cite{he2016deep}, ReDimNets~\cite{yakovlev2024reshape}, CAM++~\cite{wang2023cam++}, X-vector~\cite{snyder2018x}, MobileNetV2~\cite{sandler2018mobilenetv2}, SAM-ResNet~\cite{qin2022simple}, and Res2Net~\cite{gao2019res2net}. 
Unless otherwise stated, we used public implementations from the WeSpeaker toolkit \cite{wang2023wespeaker}. 
For MobileNetV2, we started from the official repository~\cite{sandler2018mobilenetv2} and adapted it to SV by changing the stride of the second inverted residual block from 1 to 2 and reducing the final channel dimension from 320 to 256. 
The classifier used AAM-Softmax with scale $s=32$ and margin $m=0.2$.
\vspace{-4mm}

\paragraph*{Input features and augmentation.}
Input features were 2-second log-Mel spectrograms with cepstral mean normalization. 
For ReDimNets, we used 72 Mel bins (25\,ms window, 15\,ms hop), and 80 Mel bins for the other models (25\,ms window, 10\,ms hop). 
Data augmentation included additive noise from MUSAN~\cite{snyder2015musan}, convolution with simulated room impulse responses~\cite{ko2017study}, and speed perturbation~\cite{ko2015audio} with factors of 0.9 and 1.1.
\vspace{-4mm}

\paragraph*{Training configuration.}
All models were trained for 150 epochs with a global batch size of 512 on 4$\times$A100 GPUs using the stochastic gradient descent optimizer (momentum 0.9). 
The learning rate was linearly warmed up from 0 to 0.1 over the first 6 epochs and then was exponentially decayed to 5$\times10^{-5}$.
We used attentive statistics pooling \cite{okabe2018attentive, desplanques2020ecapa}. 
Embedding dimensionality was 512 for X-vector and CAM++, and 256 for all other architectures. 
Additional training details followed the WeSpeaker pipeline \cite{wang2023wespeaker}.
\vspace{-4mm}

\paragraph*{Knowledge distillation methods and hyperparameters.}
TRKD was compared against logit-level methods: classical KD, DKD, and GKD.
We also evaluated embedding-level losses commonly used in SV---mean squared error (MSE) and cosine distance (COS)---computed between the teacher and student embeddings \cite{jung2019short, wang2019knowledge}. 
We set the softmax temperature to 4 for logit-based KD.
We used $\alpha = \lambda_{\mathcal M} = 1$ and $\beta = \lambda_{\mathcal F} = 8$ for DKD and TRKD, and GKD was implemented according to its original formulation and default settings~\cite{zhao2023grouped}. 
The cutoff value $\tau$ was decayed exponentially from 1.0 to 0.05 between epochs 10 and 60 (with $\gamma = 0.001$).

\begin{table*}[t]
\centering
\caption{
VoxCeleb1 original (\textbf{O}), extended (\textbf{E}), and hard (\textbf{H}) evaluations: EER (\%) (lower is better) for various teacher$\to$student pairs. 
Multiply--accumulate operations (MACs) are measured for a 2-second input using fvcore.
$\Delta$ (\%) is the relative improvement over the student baseline (w/o KD), averaged over O/E/H for each block. 
Abbreviations: ECAPA1024/ECAPA400: ECAPA-TDNNs with channel sizes 1024/400; RN18/34/152: ResNet-18/-34/-152; ReDim-B5/B2: ReDimNet-B5/-B2; MNV2: MobileNetV2; SAM-RN50: SAM-ResNet50; R2N34: Res2Net-34.
}
\label{tab:kd_comparison_eer}
\resizebox{\textwidth}{!}{%
\setlength{\tabcolsep}{1.4pt}
\begin{tabular}{l l
CCCC G CCCC G CCCC G CCCC G CCCC G CCCC}
\toprule
\multicolumn{2}{l}{} &
\multicolumn{15}{c}{\textit{Homogeneous}} &
\multicolumn{14}{c}{\textit{Heterogeneous}} \\
\cmidrule(lr){3-17}\cmidrule(lr){18-31}
\multicolumn{2}{l}{\textit{T $\to$ S}} &
\multicolumn{4}{@{}c@{}}{ECAPA1024 $\to$ ECAPA400} & 
& \multicolumn{4}{@{}c@{}}{RN34 $\to$ RN18} & 
& \multicolumn{4}{@{}c@{}}{ReDim-B5 $\to$ ReDim-B2} &
& \multicolumn{4}{@{}c@{}}{CAM++ $\to$ X-vector} &
& \multicolumn{4}{@{}c@{}}{RN152 $\to$ MNV2} &
& \multicolumn{4}{@{}c@{}}{SAM-RN50 $\to$ R2N34} \\
\multicolumn{2}{l}{\textit{Params (M)}} &
\multicolumn{4}{@{}c@{}}{14.265 $\to$ 4.434} & &
\multicolumn{4}{@{}c@{}}{7.292 $\to$ 4.764} & &
\multicolumn{4}{@{}c@{}}{7.651 $\to$ 4.888} & &
\multicolumn{4}{@{}c@{}}{7.439 $\to$ 4.996} & &
\multicolumn{4}{@{}c@{}}{22.447 $\to$ 3.785} & &
\multicolumn{4}{@{}c@{}}{25.214 $\to$ 6.005} \\
\multicolumn{2}{l}{\textit{MACs (G)}} &
\multicolumn{4}{@{}c@{}}{2.593 $\to$ 0.712} & &
\multicolumn{4}{@{}c@{}}{4.660 $\to$ 2.245} & &
\multicolumn{4}{@{}c@{}}{8.925 $\to$ 0.925} & &
\multicolumn{4}{@{}c@{}}{1.267 $\to$ 0.607} & &
\multicolumn{4}{@{}c@{}}{15.108 $\to$ 1.204} & &
\multicolumn{4}{@{}c@{}}{18.528 $\to$ 1.836} \\
\cmidrule(l{.4em}r{.4em}){3-6}
\cmidrule(l{.4em}r{.4em}){8-11}
\cmidrule(l{.4em}r{.4em}){13-16}
\cmidrule(l{.4em}r{.4em}){18-21}
\cmidrule(l{.4em}r{.4em}){23-26}
\cmidrule(l{.4em}r{.4em}){28-31}
\multicolumn{2}{l}{\textit{Eval set}} &
\textbf{O} & \textbf{E} & \textbf{H} & \scriptsize$\Delta$ (\%) & &
\textbf{O} & \textbf{E} & \textbf{H} & \scriptsize$\Delta$ (\%) & &
\textbf{O} & \textbf{E} & \textbf{H} & \scriptsize$\Delta$ (\%) & &
\textbf{O} & \textbf{E} & \textbf{H} & \scriptsize$\Delta$ (\%) & &
\textbf{O} & \textbf{E} & \textbf{H} & \scriptsize$\Delta$ (\%) & &
\textbf{O} & \textbf{E} & \textbf{H} & \scriptsize$\Delta$ (\%) \\
\midrule

\addlinespace[2pt]
\multirow{2}{*}{\rotatebox[origin=c]{90}{\textit{w/o KD}}} &
Teacher &
0.904 & 1.051 & 1.839 & \scriptsize -- & &
1.005 & 1.138 & 2.064 & \scriptsize -- & &
0.564 & 0.858 & 1.613 & \scriptsize -- & &
0.819 & 1.025 & 2.003 & \scriptsize -- & &
0.638 & 0.825 & 1.529 & \scriptsize -- & &
0.893 & 1.000 & 1.870 & \scriptsize -- \\
& Student &
1.351 & 1.395 & 2.607 & \scriptsize -- & &
1.462 & 1.583 & 2.817 & \scriptsize -- & &
0.867 & 1.045 & 1.973 & \scriptsize -- & &
1.899 & 1.841 & 3.355 & \scriptsize -- & &
1.479 & 1.449 & 2.603 & \scriptsize -- & &
1.383 & 1.359 & 2.422 & \scriptsize -- \\
\addlinespace[2pt]
\midrule

\multirow{2}{*}{\rotatebox[origin=c]{90}{\textit{Embd}}} &
MSE &
1.229 & 1.343 & 2.502 & \deltaup{\scriptsize +5.2} & & 
1.489 & 1.522 & 2.711 & \deltaup{\scriptsize +2.4} & &
0.920 & 1.057 & 1.984 & \deltadown{\scriptsize $-$2.0} & &
1.936 & 1.967 & 3.359 & \deltadown{\scriptsize $-$2.4} & &
1.420 & 1.488 & 2.648 & \deltadown{\scriptsize $-$0.4} & & 
1.409 & 1.393 & 2.506 & \deltadown{\scriptsize $-$2.8} \\
& COS &
1.399 & 1.363 & 2.452 & \deltaup{\scriptsize +2.6} & & 
1.505 & 1.505 & 2.664 & \deltaup{\scriptsize +3.2} & &
0.894 & 1.051 & 1.981 & \deltadown{\scriptsize $-$1.1} & & 
1.914 & 1.801 & 3.201 & \deltaup{\scriptsize +2.5} & &
1.398 & 1.433 & 2.552 & \deltaup{\scriptsize +2.7} & & 
1.255 & 1.350 & 2.441 & \deltaup{\scriptsize +2.3} \\
\midrule

\multirow{4}{*}{\rotatebox[origin=c]{90}{\textit{Logit}}} &
KD \cite{hinton2015distilling}&
1.159 & 1.241 & 2.252 & \deltaup{\scriptsize +13.1} & &
1.452 & 1.491 & 2.729 & \deltaup{\scriptsize +3.2} & &
0.776 & 1.074 & 2.033 & \deltaup{\scriptsize +0.1} & &
1.840 & 1.859 & 3.444 & \deltadown{\scriptsize $-$0.7} & &
1.127 & 1.193 & 2.271 & \deltaup{\scriptsize +17.0} & &
1.207 & 1.358 & 2.543 & \deltaup{\scriptsize +1.1} \\
& DKD \cite{zhao2022decoupled}&
1.101 & 1.200 & 2.159 & \deltaup{\scriptsize +16.7} & &
1.298 & 1.433 & 2.644 & \deltaup{\scriptsize +8.3} & &
0.729 & 0.970 & 1.826 & \deltaup{\scriptsize +9.3} & &
1.750 & 1.803 & 3.256 & \deltaup{\scriptsize +4.0} & &
1.053 & 1.184 & 2.246 & \deltaup{\scriptsize +18.9} & &
1.101 & 1.277 & 2.419 & \deltaup{\scriptsize +7.1} \\
& GKD \cite{zhao2023grouped}&
1.058 & 1.218 & 2.183 & \deltaup{\scriptsize +16.7} & &
1.324 & 1.428 & 2.628 & \deltaup{\scriptsize +8.2} & &
0.697 & 0.995 & 1.838 & \deltaup{\scriptsize +9.1} & &
1.781 & 1.791 & 3.337 & \deltaup{\scriptsize +2.6} & &
1.047 & 1.210 & 2.296 & \deltaup{\scriptsize +17.7} & &
1.138 & 1.320 & 2.463 & \deltaup{\scriptsize +4.7} \\
\rowcolor{TRKDhl}
& \textbf{TRKD} &
\textbf{0.978} & \textbf{1.115} & \textbf{2.001} & \deltaup{\scriptsize \textbf{+23.5}} & &
\textbf{1.212} & \textbf{1.322} & \textbf{2.461} & \deltaup{\scriptsize \textbf{+14.8}} & &
\textbf{0.627} & \textbf{0.885} & \textbf{1.644} & \deltaup{\scriptsize \textbf{+18.8}} & &
\textbf{1.595} & \textbf{1.692} & \textbf{3.164} & \deltaup{\scriptsize \textbf{+9.1}} & &
\textbf{0.883} & \textbf{1.068} & \textbf{2.016} & \deltaup{\scriptsize \textbf{+28.3}} & &
\textbf{0.968} & \textbf{1.157} & \textbf{2.178} & \deltaup{\scriptsize \textbf{+16.7}} \\
\bottomrule
\end{tabular}
\vspace{-5mm}
}
\end{table*}

\section{Results}
We validated the proposed method on VoxCeleb1-O/E/H, comparing a broad set of distillation approaches across six different teacher--student pairings (Table~\ref{tab:kd_comparison_eer}). 
The experiments covered both homogeneous and heterogeneous transfer scenarios to assess any architecture-dependent effects. 
Compared to the student baseline (w/o KD), embedding-level KD (MSE, COS) yielded limited, inconsistent gains and even degraded EER, indicating that matching embeddings alone is insufficient.
By contrast, logit-level KD (classical KD, DKD, GKD) delivered consistent improvements by aligning the class posterior distributions, which are directly optimized by the classifier. 
TRKD attained the lowest EER on all 18 evaluations, outperforming the strongest prior logit-level method in every teacher--student pairing. 
Averaged over all conditions, TRKD reduced EER by 18.7\% relative to the student baseline (w/o KD), with particularly large gains for lightweight students and cross-architecture transfers.
These results suggest that TRKD suppresses long-tail noise and concentrates supervision on confusable classes, enabling more reliable knowledge distillation. 

\begin{figure}[!t]
    \centering
    \includegraphics[width=\linewidth]{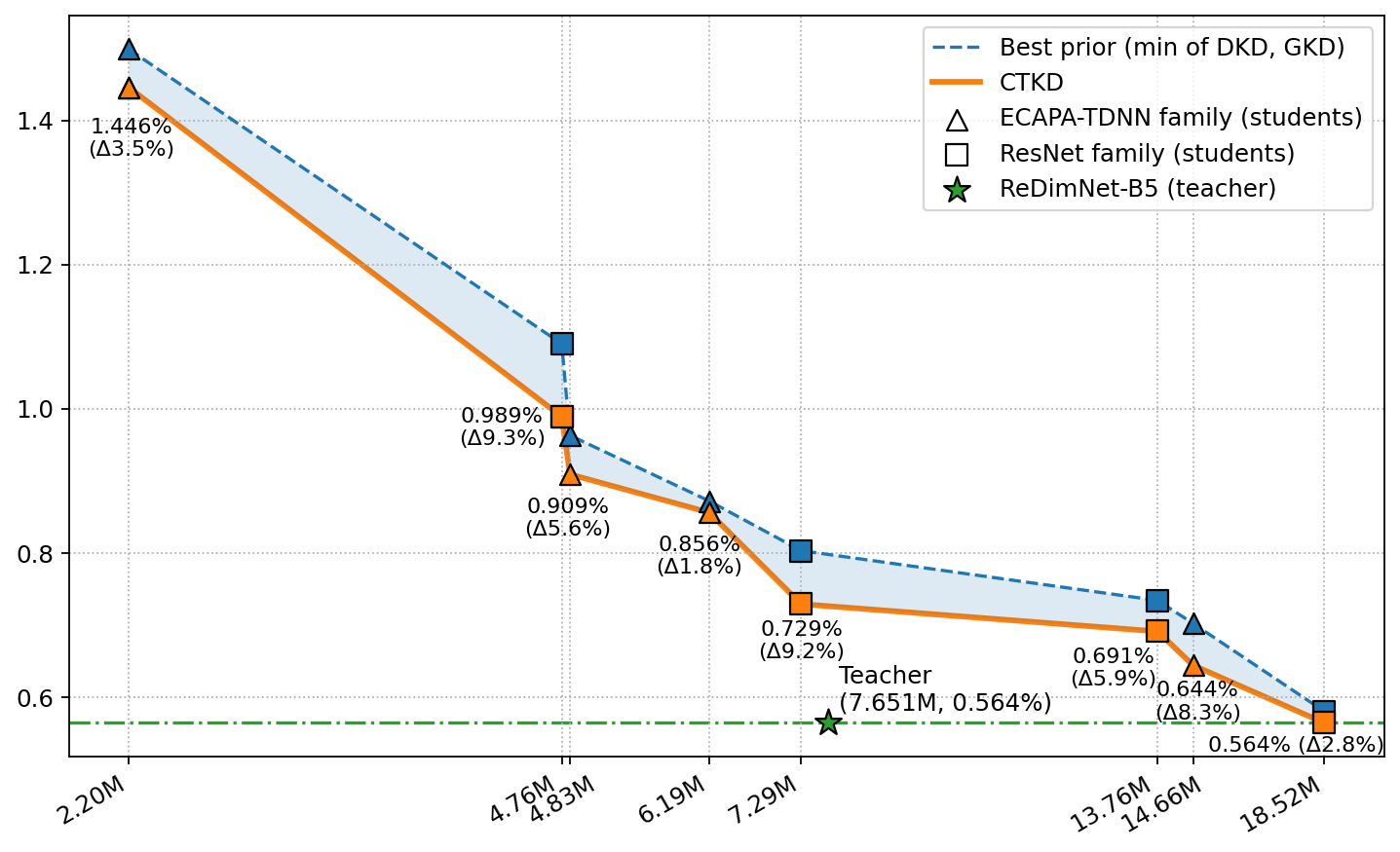}
    \vspace{-5mm}
    \caption{
    EER (\%) versus student parameter count (log scale), comparing TRKD to the best prior approach (the minimum of DKD and GKD) across different student model sizes. 
    The teacher is fixed at ReDimNet-B5 (dash--dot horizontal line; star marker). 
    Students are drawn from two architecture families: ECAPA-TDNN (triangle markers) and ResNet (square markers), ordered left-to-right by increasing parameter count (ECAPA128$\to$ECAPA400$\to$ECAPA512$\to$ECAPA1024; RN18$\to$RN34$\to$RN50$\to$RN101).  
    Labels on the TRKD curve (solid line) indicate each student’s EER and its relative improvement $\Delta$ (\%) over the best prior (dashed line); the shaded region denotes this gain.
    }
    \label{fig:visualization}
    \vspace{-1mm}
\end{figure}

Fig.~\ref{fig:visualization} plots EER on the y-axis versus student model size, with the x-axis on a log scale and the teacher fixed at ReDimNet-B5. 
By fixing the teacher, we isolated the effects of student capacity and architecture. 
The student pool comprises ECAPA-TDNNs and ResNets. 
Each marker is labeled with the TRKD EER and its relative improvement $\Delta$ (\%) over the best prior logit method (the minimum of DKD and GKD). 
Consistent gains were observed across student model sizes---typically 2--9\% (macro-average $\approx 5.8\%$)---peaking for mid-sized students (RN18, RN34) and remaining non-trivial even when the student is larger than the teacher (ECAPA1024, RN101). 
Notably, TRKD brought the RN101 student (18.5M parameters) down to the teacher’s EER (0.564\%), whereas the best prior remained higher. 
Overall, TRKD reduced EER across model scales and further narrowed the teacher--student performance gap. 

\begin{table}[t]
\centering
\caption{Ablation of TRKD components on VoxCeleb1-O (ReDimNet-B5$\to$ReDimNet-B2). 
$\Delta$ (\%) is the relative improvement over DKD, and $\tau$ is the cumulative cutoff used to form the confusion-set.}
\label{tab:TRKD_ablation}
\begin{threeparttable}
\setlength{\tabcolsep}{4.0pt}
\begin{tabular}{l l c c c}
\toprule
\textbf{ID} & \textbf{Method (replacement)} & \textbf{$\tau$ schedule} & \textbf{EER (\%)} & \textbf{$\Delta$ (\%)} \\
\midrule
\#1 & DKD ($\mathcal{L}_{\mathrm{TCKD}}$ + $\mathcal{L}_{\mathrm{NCKD}}$)                     & fixed 1.0\tnote{*}        & 0.729 & --   \\
\addlinespace[2pt]
\multicolumn{5}{l}{\emph{Replace a single DKD term}} \\
\#2 & $\mathcal{L}_{\mathrm{TCKD}}$ $\to$ $\mathcal{L}_{\mathrm{TMKD}}$             & fixed 0.05       & \multicolumn{1}{c}{\textit{diverged}} & -- \\
\#3 & $\mathcal{L}_{\mathrm{TCKD}}$ $\to$ $\mathcal{L}_{\mathrm{TMKD}}$             & 1.0 $\to$ 0.05   & 0.691 & +5.2 \\
\#4 & $\mathcal{L}_{\mathrm{NCKD}}$ $\to$ $\mathcal{L}_{\mathrm{CFKD}}$     & 1.0 $\to$ 0.05   & 0.654 & +10.3 \\
\addlinespace[2pt]
\#5 & \textbf{TRKD ($\mathcal{L}_{\mathrm{TMKD}}$ + $\mathcal{L}_{\mathrm{CFKD}}$)}                      & 1.0 $\to$ 0.05   & \textbf{0.627} & \textbf{+14.0} \\
\bottomrule
\end{tabular}
\begin{tablenotes}\footnotesize
\item[*] DKD is algebraically equivalent to TRKD with a fixed $\tau=1.0$
\end{tablenotes}
\end{threeparttable}
\vspace{-2mm}
\end{table}

Table~\ref{tab:TRKD_ablation} presents an ablation of TRKD on VoxCeleb1-O (ReDimNet-B5$\to$ReDimNet-B2). 
The DKD baseline (ID~\#1) yielded an EER of 0.729\%. 
As noted, DKD is algebraically equivalent to TRKD when $\tau$ is fixed at 1.0: the confusion-set coincides with all non-target classes (no background-set), and $\mathcal{L}_{\mathrm{TMKD}}=\mathcal{L}_{\mathrm{TCKD}}$ while $\mathcal{L}_{\mathrm{CFKD}}=\mathcal{L}_{\mathrm{NCKD}}$. 
Replacing $\mathcal{L}_{\mathrm{TCKD}}$ with $\mathcal{L}_{\mathrm{TMKD}}$ (ID~\#2) caused training to diverge when $\tau$ was fixed at 0.05, but training became stable and the EER improved to 0.691\% (+5.2\,\%) under a curriculum on $\tau$ (1.0$\to$0.05; ID~\#3), thereby transferring inter-class confusion as well as difficulty. 
Replacing $\mathcal{L}_{\mathrm{NCKD}}$ with $\mathcal{L}_{\mathrm{CFKD}}$ (ID~\#4) under the same curriculum further reduced the EER to 0.654\,\% (+10.3\,\%), indicating that the confusion-set conditional term contributed more than the three-mass term alone. 
Combining both replacements (TRKD, ID~\#5) achieved the best result of 0.627\% EER (+14.0\,\%). 
In summary, the ablation results showed that (i) a curriculum on $\tau$ was crucial for stability, and (ii) $\mathcal{L}_{\mathrm{TMKD}}$ and $\mathcal{L}_{\mathrm{CFKD}}$ were complementary, jointly yielding the largest gain.

\section{Conclusion}
We introduced triage knowledge distillation (TRKD) for speaker verification (SV), which combines a coarse three-mass alignment with a fine-grained confusion-set conditional while ignoring the background-set. 
By scheduling a cumulative-probability cutoff $\tau$ over non-target classes, TRKD stabilizes training and concentrates supervision on the most confusable impostors, suppressing long-tail noise. 
On VoxCeleb1-O/E/H, TRKD attained the lowest EER across all teacher--student pairings and achieved an average relative improvement of 18.7\% compared to students trained without KD. 
Ablation studies confirm that (i) a $\tau$-curriculum is critical for stability and (ii) the three-mass and confusion-set terms provide complementary gains. 
Future work includes extending TRKD to variable-length and far-field SVs and exploring its applicability to other domains such as computer vision and natural language processing.

\bibliographystyle{IEEEbib}
\bibliography{myrefs}
\end{document}